\documentclass[reprint,amsmath,amssymb,aps,superscriptaddress,prstab]{revtex4-2}
\usepackage{graphicx}
\usepackage{dcolumn}
\usepackage{bm}
\usepackage{siunitx}
\usepackage{color}
\usepackage{amsmath}
\usepackage{url} 
\usepackage{changes}

\usepackage[colorlinks, citecolor=blue, urlcolor=blue, bookmarks=false, linkcolor=blue, hypertexnames=true]{hyperref}

\begin{document}

\preprint{APS/123-QED}

\title{Role of On-site and Inter-site Coulomb Interactions in KV$_3$Sb$_5$: A first-principles DFT+$U$+$V$ study\\
}

\author{Indukuru Ramesh Reddy}
\thanks{These authors contributed equally to this work}
\affiliation{Department of Physics, Kyungpook National University, Daegu 41566, Republic of Korea}
\affiliation{Department of Physics, Kunsan National University, Gunsan 54150, Republic of Korea}
\author{Sayandeep Ghosh}
\thanks{These authors contributed equally to this work}
\affiliation{Department of Physics, Kunsan National University, Gunsan 54150, Republic of Korea}
\affiliation{Department of Physics, Chungnam National University, Daejeon 34134, Republic of Korea}
\author{Bongjae Kim}
\email{bongjae@knu.ac.kr}
\affiliation{Department of Physics, Kyungpook National University, Daegu 41566, Republic of Korea}
\affiliation{Department of Physics, Kunsan National University, Gunsan 54150, Republic of Korea}
\author{Chang-Jong Kang}
\email{cjkang87@cnu.ac.kr}
\affiliation{Department of Physics, Chungnam National University, Daejeon 34134, Republic of Korea}
\affiliation{Institute of Quantum Systems, Chungnam National University, Daejeon 34134, Republic of Korea}

\date{\today}
\begin{abstract}

Nonlocal Coulomb interactions play a crucial role in stabilizing distinct electronic phases in kagome materials. In this work, we systematically investigate the effects of on-site ($U$) and inter-site ($V$) Coulomb interactions on the electronic structure and stability of charge-density-wave (CDW) phases in the kagome metal KV$_3$Sb$_5$ using density functional theory (DFT+$U$+$V$) calculations. We demonstrate that
$V$ promotes the formation and stability of CDW phases, whereas
$U$ suppresses these phases, highlighting a fundamental competition between local and nonlocal 
Coulomb interactions. By directly comparing our theoretical results with angle-resolved photoemission spectroscopy (ARPES) data, we identify realistic values of $U$ and $V$ that accurately describe the electronic band structure of KV$_3$Sb$_5$. Our findings establish a detailed $U$–$V$ phase diagram for KV$_3$Sb$_5$, offering valuable insights into the correlated electronic states in kagome metals and serving as a foundation for future explorations of correlation-driven phenomena in related materials.

\end{abstract}

\maketitle

\section{Introduction}

Kagome lattices are formed by a periodic arrangement of corner-sharing triangles, resulting in a distinctive crystal structure with inherent geometric frustration \cite{freedman2010site}. This geometric frustration arises from the nonequivalence of the three vertices in each triangular unit, which prevents the system from achieving long-range magnetic ordering and instead stabilizes disordered spin states. Consequently, kagome lattices with intrinsic magnetic frustration are promising candidates for hosting quantum spin liquid states \cite{balents2010QSL, norman2016QSL, georgopoulou2023QSL, han2012fractionalized}.

In addition, kagome lattices exhibit topologically nontrivial electronic states linked to local electronic symmetry breaking~\cite{ye2018massive, yin2018giant, liu2020orbital, kang2020dirac, morali2019fermi}. Their complex band structures—featuring flat bands, Dirac cones, and van Hove singularities—heighten their significance in condensed matter physics \cite{yin2022topological}. The interplay between strong Coulomb interactions and nontrivial band topology, driven by the diverging density of states at van Hove singularities and flat bands, gives rise to diverse exotic phenomena. These include topological insulator \cite{guo2009topological}, superconductivity \cite{ko2009doped}, charge density wave (CDW) \cite{wang2013competing}, charge fractionalization \cite{ruegg2011fractionally,tomonari2017cf}, and Dirac/Weyl semimetals \cite{morali2019fermi,liu2019magnetic}.

\begin{figure}[!!t]
\begin{center}
 \includegraphics[trim=3.5cm 3.5cm 2.4cm 2cm, clip=true,scale=0.45]{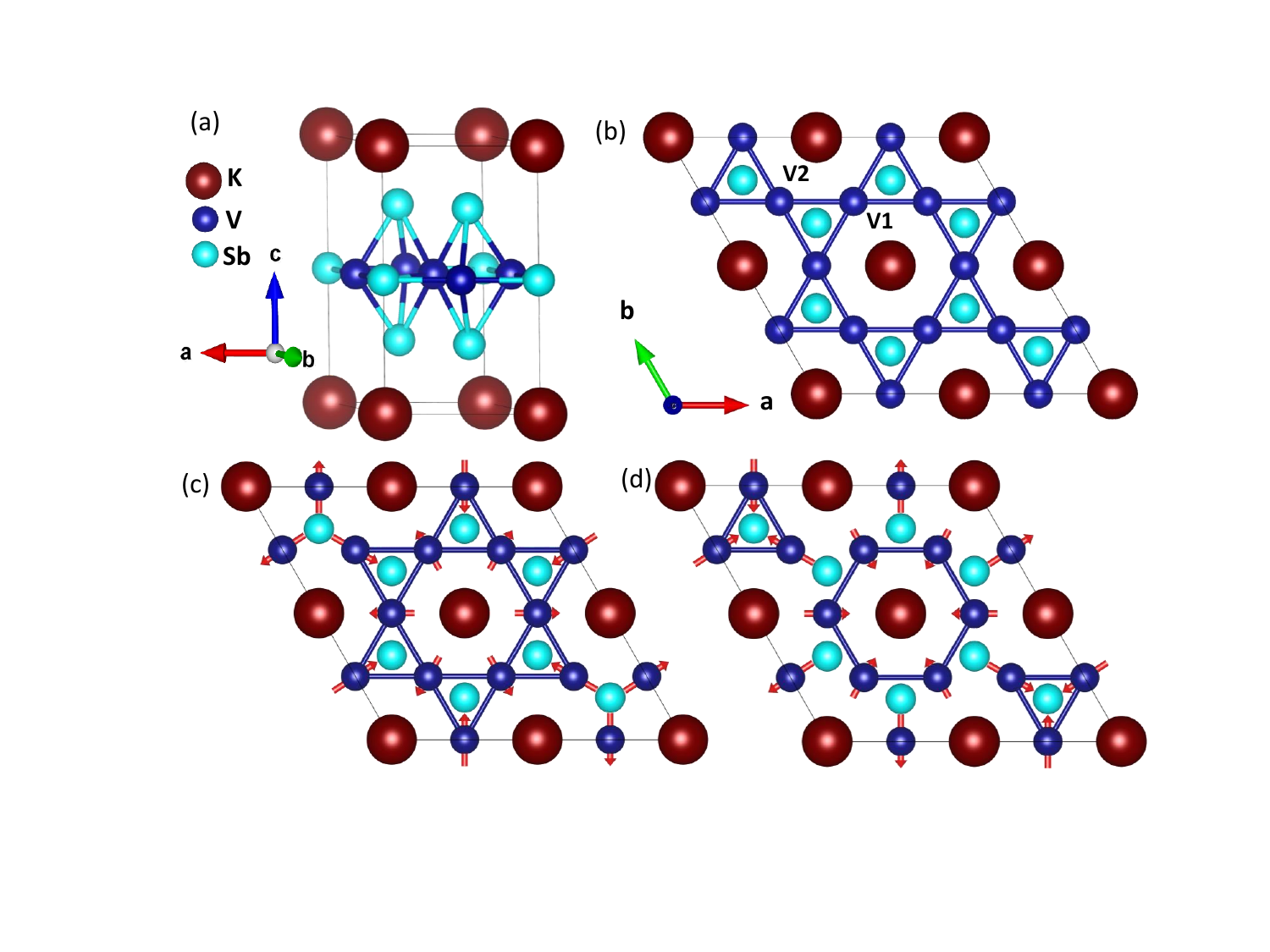}
\end{center}
\caption{(a) Unit cell and (b) $2\times2\times1$ supercell crystal structure of KV$_3$Sb$_5$, where the vanadium atoms exhibit kagome net and the V-Sb layers are intercalated by a layer of K atoms. (c) Star of David (SD) and (d) inverse Star of David (ISD) $2\times2$ charge order distortion in KV$_3$Sb$_5$. The K, V, Sb atoms are represented by brown, dark blue, and light blue balls, respectively. V1 and V2 represent the inequivalent vanadium sites of the kagome lattice.
}
\label{fig:schematic}
\end{figure}

The recently discovered non-magnetic vanadium-based kagome lattice system
{\it A}V$_3$Sb$_5$ ({\it A} = K, Rb, Cs) ~\cite{ortiz2019new}
has received significant attention due to its fascinating electronic phenomena,
making it one of the most intensively studied kagome systems \cite{neupert2022charge,wang2021charge,yin2022topological}.
In {\it A}V$_3$Sb$_5$,
the vanadium atoms crystallize in a uniform kagome lattice
within the V$_3$Sb$_5$ layers sandwiched between $A$ layers
as shown in Fig. \ref{fig:schematic}.
All three compounds, KV$_3$Sb$_5$, RbV$_3$Sb$_5$ and CsV$_3$Sb$_5$, undergo an intriguing CDW transition with T$_{\text{CDW}}$ =78 K, 102 K, and 94 K.
This transition is associated with in-plane $2\times2$ periodicity
\cite{jiang2021unconventional,zhao2021cascade,liang2021three,chen2021roton}
and is accompanied by additional out-of-plane doubling ($2\times2\times2$)~\cite{liang2021three,li2021observation}
or quadrupling ($2\times2\times4$)~\cite{ortiz2021fermi}.
In addition, these compounds exhibit superconductivity at temperatures
ranging from
0.9 K (KV$_3$Sb$_5$, RbV$_3$Sb$_5$) to 2.5 K
(CsV$_3$Sb$_5$)~\cite{ortiz2019new,ortiz2020cs,ortiz2021superconductivity}.
The interplay among CDW, superconductivity, lattice frustration,
and band topology
has led to the discovery of interesting electronic properties,
including a strong anomalous Hall effect~\cite{yang2020giant,yu2021concurrence},
time-reversal-symmetry-breaking CDW and
charge order states
\cite{mielke2022time,jiang2021unconventional,xu2022three,guguchia2023tunable},
pair density wave \cite{chen2021roton},
and rotational-symmetry-breaking electronic nematicity
\cite{xu2022three,nie2022charge,wang2025discovery}.

Regarding their crystal structures, initial results from X-ray diffraction~\cite{ortiz2020cs,ortiz2021superconductivity}
and scanning tunneling microscopy (STM)
\cite{jiang2021unconventional,zhao2021cascade,liang2021three}
indicate
the presence of a $2\times2$ CDW phase.
A recent density functional theory (DFT) study, in comparison with STM data,
confirmed the inverse Star of David (ISD) distortion
as the ground state for {\it A}V$_3$Sb$_5$ compounds~\cite{tan2021charge}.
Subsequent studies have provided further evidence supporting the ISD deformation
in these systems \cite{kato2022three,kang2022twofold}.

Important features of the electronic structures in $A$V$_3$Sb$_5$ compounds are the van Hove singularities near the Fermi energy, which have V-$3d$ orbital character.
The precise position of these van Hove singularities depends on details of the Coulomb interactions, inducing electronic instabilities that lead to the emergence of various CDW phases~\cite{jiang2021unconventional,denner2021analysis,jeong2022crucial}.
In particular, the chiral CDW phase observed in $A$V$_3$Sb$_5$ has been plausibly attributed to \textit{nonlocal} Coulomb interactions~\cite{denner2021analysis}.
Other contributing factors, such as lattice deformations, electron-phonon coupling, and Fermi surface nesting, can also interplay with electronic correlations and play significant roles in mediating the CDW phase transition
\cite{tan2021charge,wenzel2022optical,xie2022electron,ratcliff2021coherent,luo2022electronic,wu2022charge,liu2022observation}.

The impact of Coulomb interactions has been extensively investigated using
Kagome Hubbard model calculations, primarily within single-band models.
Kiesel \textit{et al.} \cite{PhysRevB.86.121105,PhysRevLett.110.126405}
explored the system
using a single-band extended Hubbard model
that incorporates 
both on-site ($U$) and inter-site ($V$) Coulomb interactions. 
They studied various phases arising from Fermi surface instabilities
within the $U$-$V$ phase diagram \cite{PhysRevB.86.121105,PhysRevLett.110.126405}. 
Notably, near the van Hove filling, 
$V$ can promote distinct CDW phases, which explain several key experimental findings~\cite{ferrari2022charge}.
The presence of sizable long-range interactions has been also reported,
which exhibit interesting universality across kagome metals~\cite{PhysRevResearch.5.L012008}.
More recent studies have further emphasized the critical role of inter-site Coulomb interactions
in understanding the electronic structure of kagome materials,
investigating their impact on a broad range of electronic properties
\cite{wu2021nature,zheng2022emergent,kiran2024effect,jiang2025fege,dong2023loop,bigi2024pomeranchuk,wenger2024theory}.

In this study, we perform DFT calculations,
incorporating both $U$ and $V$ Coulomb interactions to explore their specific roles in the electronic structure of KV$_3$Sb$_5$.
We examine the energy profiles associated with in-plane distortions,
leading to the Star of David (SD) and ISD phases,
along with the variations in $U$ and $V$.
We also explore the effects of $U$ and $V$ on different stacking arrangements
along the out-of-plane direction of various CDW phases
and present the resulting $U$-$V$ phase diagram for KV$_3$Sb$_5$.
A comparison with the experimental ARPES data and our calculations enables us to locate the position of the system in the phase space.

\begin{figure*}[!!ht]
\begin{center}
\includegraphics[trim=0cm 5.2cm 0cm 3.5cm, clip=true,scale=0.7]{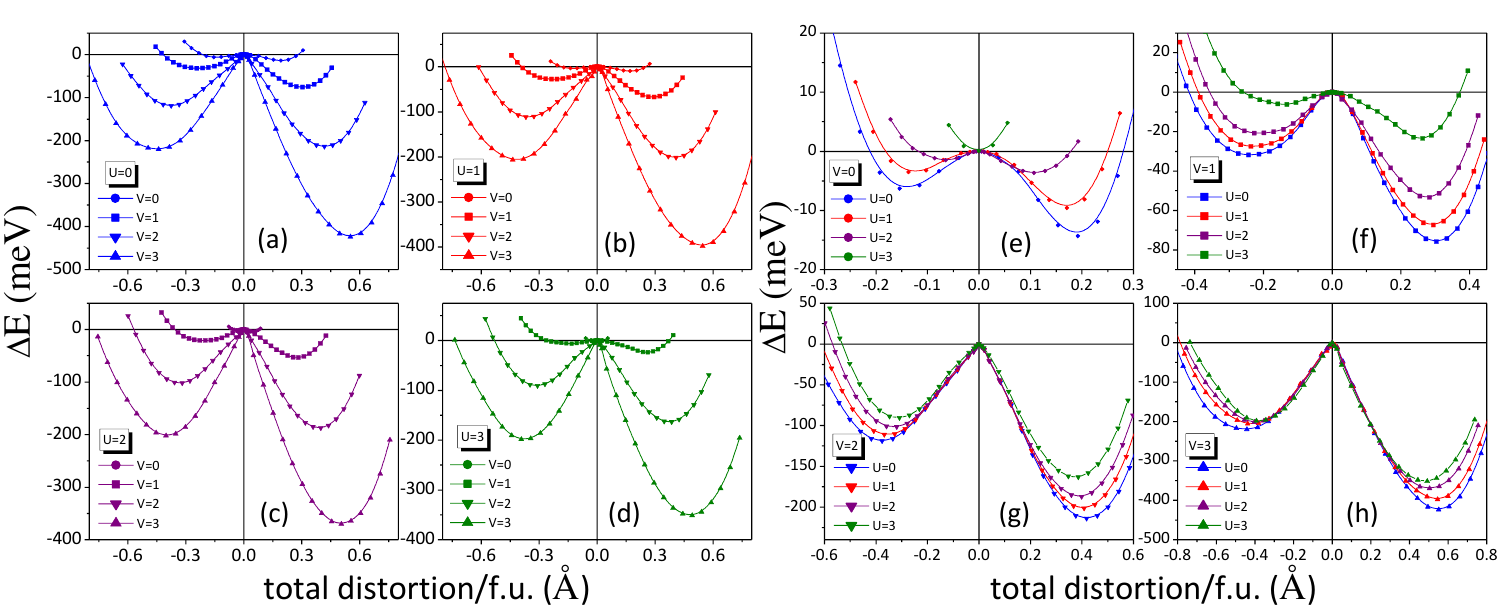}
\end{center}
\caption{(a) Variation of energy profiles as a function of the total distortions per formula unit (f.u.) for KV$_3$Sb$_5$: (a-d) for fixed $U = $ 0, 1, 2, and 3 eV with varying $V$,
and (e-h) for fixed $V =$ 0, 1, 2, and 3 eV with varying $U$.
The total distortion refers to the sum of the displacements of all vanadium atoms
within the $2\times2\times1$ supercell.
$\Delta E$ represents the energy difference relative to the pristine phase.
The positive (negative) distortion corresponds to the magnitude of the breathing (shrinking) phonon mode of the $2\times2$ inverse Star of David (Star of David) CDW phase.}

\label{fig:energy}
\end{figure*} 

\section{Computational details}

Density functional theory (DFT) calculations were performed using the plane-wave pseudopotential method, as implemented in the Quantum Espresso package~\cite{giannozzi2009quantum, giannozzi2017advanced, giannozzi2020quantum}.
The exchange-correlation functional was treated using the generalized gradient approximation (GGA) parameterized by Perdew-Burke-Ernzerhof (PBE)~\cite{perdew1996generalized}.
Pseudopotentials were obtained from the Standard Solid-State Pseudopotentials (SSSP) library, version 1.2.1~\cite{SSSP_lib}.
We considered a $2\times2\times1$ supercell of KV$_3$Sb$_5$, containing 36 atoms,
to incorporate two relevant CDW phases$-$SD and ISD$-$on equal footing (see Fig.~\ref{fig:schematic}),
and performed all calculations using a $5\times5\times5$ $k$-mesh.
The plane-wave energy cut-off was set to 37 Ry.
For density of states calculations, a denser $8\times8\times8$ $k$-mesh was used.
Band unfolding calculations were carried out using the unfold.x code~\cite{pacile2021narrowing}, based on the method of Popescu and Zunger~\cite{popescu2012extracting}.
Spin-orbit coupling and van der Waals corrections were not included,
as they do not affect the main conclusions of this study.

The optimized structure obtained from the DFT calculations yields
lattice parameters of $a = 5.48$ \AA~and $c = 9.35$ \AA.
These optimized lattice parameters are in good agreement with experimental data~\cite{ortiz2019new,ortiz2021superconductivity}
and previous DFT studies~\cite{tan2021charge,ortiz2021superconductivity}.
Neglecting van der Waals corrections results in a slightly larger $c$ lattice parameter compared to experimental reports ($a_{\text{expt}}=5.48$ \AA, $c_{\text{expt}}= 8.958$ \AA)~\cite{ortiz2021superconductivity,ortiz2019new}.
Details of the structural parameters are provided in Table S1 of the Supplementary Material~\cite{Supp}.

The Coulomb interaction parameters ($U$ and $V$) for the V-$3d$ manifold were calculated using the constrained random-phase approximation (cRPA) with the projector method \cite{kaltak2015merging} implemented in VASP \cite{kresse1996efficient}.
A $8\times8\times5$ $k$-mesh and 288 bands were employed for the cRPA calculations.
To construct the Wannier functions, all V-$3d$ and Sb-$5p$ orbitals were considered, with a disentanglement energy window of [-5.0, 4.8] eV (see Fig. S1).
The Wannier functions were obtained using WANNIER90 \cite{mostofi2008wannier90} and the VASP2WANNIER \cite{franchini2012maximally} interface,
with default parameters \cite{reddy2024exploring} for disentanglement and Wannierization.
A large plane-wave cutoff energy of 500 eV was used for self-consistent calculations, while 400 eV was used for dielectric function calculations.
The resulting band structures and interaction parameters are provided in the Supplementary Material \cite{Supp}.
Using the $U$ and $V$ values obtained from cRPA calculations,
we employed the DFT+$U$+$V$ method,
as implemented in Quantum Espresso \cite{campo2010extended},
to investigate the electronic structure of KV$_3$Sb$_5$.
In addition, we systematically varied the values of $U$ and $V$ to further investigate their impact on the ground state and electronic structure of KV$_3$Sb$_5$.

\section{Results and Discussion}
\patchcmd{\subsection}
  {\centering}
  {\raggedright}
  {}
  {}

\subsection{Pristine and CDW structures}
\label{sub:struct}
Figures~\ref{fig:schematic}(a) and (b) show the unit cell and $2\times2$ supercell of KV$_3$Sb$_5$, respectively, with space group P6/$mmm$ (No. 191).
The supercell contains two inequivalent vanadium sites, denoted as V1 and V2,
whereas the unit cell contains only one.
The distinct Wyckoff positions for V1 and V2 are provided in Table S1 of the Supplementary Material~\cite{Supp}.
The vanadium sublattice forms a perfect kagome lattice structure, with the V-Sb layers inserted between the K atoms. Two different $2\times2$ CDW structures, SD (Fig. \ref{fig:schematic}(c)) and ISD (Fig. \ref{fig:schematic}(d)), were constructed by incorporating opposite lattice distortions on the V atoms relative to the pristine supercell.

\subsection{Stability of CDW phases depending on $U$ and $V$}
\label{sub:stability}

First, we discuss the stability of the CDW phases upon introducing $U$ and $V$
in the kagome metal KV$_3$Sb$_5$.
Figure~\ref{fig:energy} shows the energy profiles
as a function of vanadium atomic distortions for different values of $U$ and $V$. 
Positive (negative) distortion values correspond to the ISD (SD) phase.
Our results clearly demonstrate that,
in the absence of $U$ and $V$,
the ISD deformation is energetically more favorable than the SD deformation.
This is consistent with experimental STM images and previous DFT results,
where the ISD phase is found to be more stable than the SD phase~\cite{tan2021charge,jiang2021unconventional}.
For all fixed values of $U$,
increasing $V$ not only enhances the stability of both the ISD and SD phases
but also increases atomic distortions  (see Figs.~\ref{fig:energy}(a-d)).
The ISD phase remains more stable than the SD phase upon variation of $V$.
For instance, at $U$ = 0, 
the atomic distortion increases by a factor of 2.9
(ISD phase) and 3.2 (SD phase) for $V=3$ eV compared to the case of $V=0$.
Furthermore, the energy difference between the ISD and SD phases
reaches 204 meV, a significant increase from 8 meV at $U$ = 0 and $V=0$.

Interestingly, for a fixed $V$, increasing $U$ progressively reduces the stability of both the ISD and SD phases. Eventually, at $V = 0$ and $U = 3$ eV, the pristine phase becomes energetically more favorable than the CDW phases (see Fig.~\ref{fig:energy}(e)). This demonstrates that strong local Coulomb interaction alone \textit{cannot} stabilize either of the CDW phases. For finite values of $V$, higher $U$ values ($\geq 3$ eV) are required to stabilize the pristine phase relative to the CDW phases. These results indicate that $V$ and $U$ exert opposite effects: while $V$ stabilizes CDW phases, $U$ suppresses them. This contrast provides compelling evidence that incorporating both $U$ and $V$ leads to fundamentally different physical consequences in the kagome metal KV$_3$Sb$_5$.

We compare these findings for KV$_3$Sb$_5$ with those for another kagome metal, CsV$_3$Sb$_5$.
Employing the similar approaches,
we obtained the corresponding energy profile for CsV$_3$Sb$_5$ as shown in Fig. S2 of the Supplementary Material~\cite{Supp}.
For a fixed $U$, CsV$_3$Sb$_5$ follows a similar trend to KV$_3$Sb$_5$, except when $V=0$.
When $V=0$, as $U$ increases (for $U < 3$ eV), the ISD phase gradually destabilizes prior to the SD phase.
At $U=3$ eV, the SD phase also becomes destabilized, approaching the pristine phase.
These results shows common tendency toward ISD phase for both systems. Note that the effective values of $U$ and $V$ in CsV$_3$Sb$_5$
are smaller than in KV$_3$Sb$_5$ suggesting the stronger CDW instability of the system.

\begin{figure}[!!t]
\begin{center}
\centering
\includegraphics[clip=true,scale=0.5]{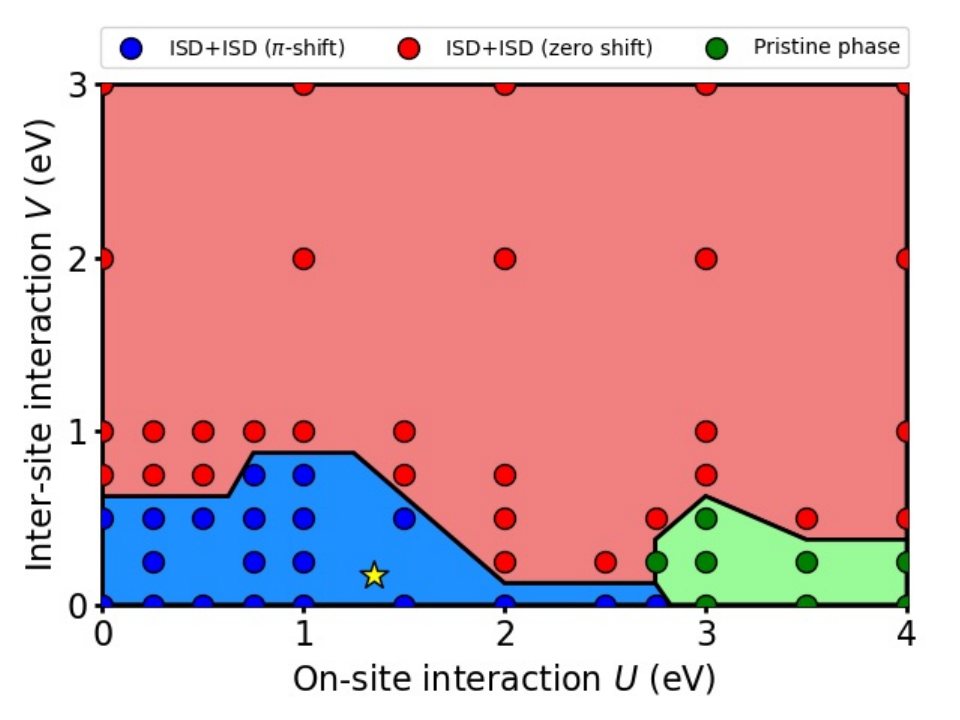}
\end{center}
\caption{Phase diagram of KV$_3$Sb$_5$ of depending on $U$ and $V$. The symbols indicate the values of $U$ and $V$ for which the calculations have been performed.
The filled blue, red, and green circle symbols represent the ISD+ISD ($\pi$-shift), ISD+ISD (zero-shift), and pristine phase, respectively. The yellow star symbol represents the $U$ and $V$ values obtained from cRPA.}
\label{fig:phase-diagram}
\end{figure} 
 
\subsection{Phase diagram of KV$_3$Sb$_5$}

Previous studies on the $A$V$_3$Sb$_5$ system have demonstrated
the possibility of various CDW phases,
which not only differentiate in-plane SD and ISD configurations
but also include phase shifts between adjacent kagome layers
\cite{liang2021three,li2021observation,ortiz2021fermi,tan2021charge}
(see Fig. S3).
Motivated by these findings, we considered six distinct CDW phases
by employing the SD and ISD deformations with either zero or $\pi$ phase shifts
between adjacent kagome layers in a $2\times2\times2$ supercell.
These six CDW phases include (a) ISD+ISD ($\pi$-shift), (b) ISD+ISD (zero-shift), (c) SD+SD ($\pi$-shift), (d) SD+SD (zero-shift), (e) SD+ISD (zero-shift), and (f) SD+ISD ($\pi$-shift). Note that the ISD+ISD with zero phase shift is equivalent to the $2\times2\times1$ ISD structure discussed in Sec.~\ref{sub:stability}.

Figure~\ref{fig:phase-diagram} shows the overall CDW phase diagram of KV$_{3}$Sb$_{5}$
as a function of $U$ and $V$.
We first note that within the studied $U$-$V$ parameter range,
the ISD deformation is consistently favored, as shown in Fig.~\ref{fig:energy}.
This preference arises because the energy scale of the inter-layer coupling is significantly smaller than that of the intra-layer CDW distortion.
Except in the limit of large $U$, where the pristine phase is stabilized for small $V$,
ISD-related phases remain more stable than SD phases throughout the phase diagram.

Among the various phases, two dominant ISD-type CDW phases,
ISD+ISD ($\pi$-shift) and ISD+ISD (zero-shift), are noted in the phase diagram.
We found that in the small $U$ and $V$ regime, ISD+ISD ($\pi$-shift) is stabilized,
as shown in Fig.~\ref{fig:phase-diagram}.
We find that inter-site Coulomb interactions
promote coherent ISD deformations across different kagome layers,
favoring a zero-shift phase alignment.
For sufficiently large $V$,
ISD+ISD (zero-shift) is always stabilized regardless of $U$.
However, in the small $V$ regime, competition arises
between ISD+ISD ($\pi$-shift) and either ISD+ISD (zero-shift) or the pristine phase,
depending on whether $U$ is small or large.
This finding is intriguing because
our Coulomb interaction profiles are restricted to the in-plane interactions,
yet they selectively stabilize specific inter-layer phase shifts.
It also highlights that
competing phases can be stabilized by subtle details of the electronic structure.

Previous studies have identified the ISD+ISD ($\pi$-shift) phase
as the ground state of the $A$V$_3$Sb$_5$ system \cite{liang2021three,jin2024pi,li2022pi}.
Hence, we can determine the realistic regime of the on-site $U$
and inter-site $V$ Hubbard parameters for the KV$_3$Sb$_5$ system
in the $U$-$V$ phase diagram:
small to moderate $U$ values and small $V$ values
(see the blue-filled areas in Fig.~\ref{fig:phase-diagram}).
To further quantify the Coumonb interaction parameters,
we performed cRPA calculations, obtaining $U$ = 1.35 eV and $V$ = 0.17 eV.
The corresponding point is marked with a star symbol in Fig.~\ref{fig:phase-diagram},
demonstrating that KV$_3$Sb$_5$ lies well within the ISD+ISD ($\pi$-shift) regime,
consistent with experimental findings.
The validity of these $U$ and $V$ values obtained from cRPA
will be further examined in the next section through
direct comparison between the band dispersion from DFT+$U$+$V$ calculations
and experimental ARPES data.

\subsection{Electronic structures: band structure}
\label{sub:band-unfolding}
\begin{figure}[!!t]
	\begin{center}
		\includegraphics[clip=true,scale=0.76]{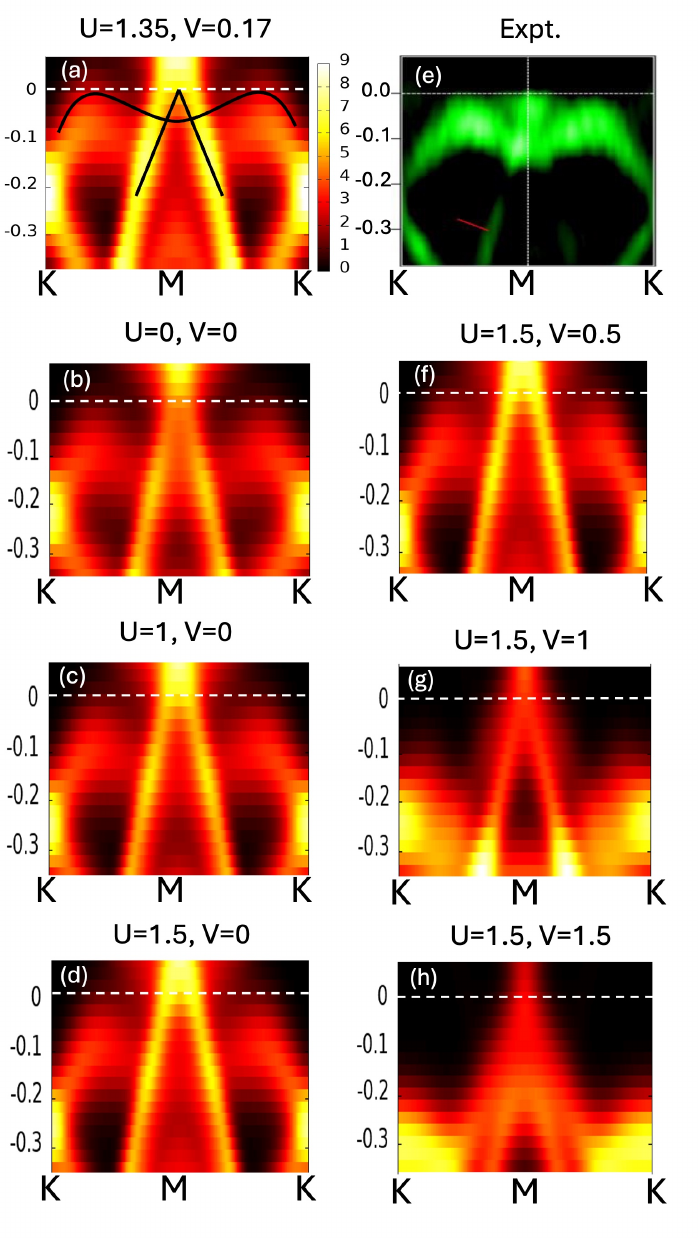}
	\end{center}
	\caption{Calculated unfolded band dispersion for $2\times2$ ISD phase along K-M-K path for different values of $U$ and $V$ (in eV): (a) cRPA values ($U=1.35, V=0.17$), (b) $U=0$, $V=0$, (c) $U=1$, $V=0$, (d) $U=1.5$, $V=0$, (f) $U=1.5$, $V=0.5$, (g) $U=1.5$, $V=1$, and (h) $U=1.5$, $V=1.5$.
    (e) ARPES intensity measured at $T$ = 20 K along the K-M-K cut at $h\nu = 114$ eV, as adapted from Kato \emph{et al.}~\cite{kato2022three}. The black solid lines in (a) are a guide for the eye to understand both the SP1 and SP2 bands.
    A Fermi level offset of 200 meV is introduced in the DFT+$U$+$V$ bands
    to enable direct comparison with the experimental spectra.}
	\label{fig:bands}
\end{figure}

We now investigate the electronic structure of KV$_3$S$_5$,
focusing on its dependence on the Hubbard parameters $U$ and $V$.
Given the weak interlayer interactions,
we consider the $2\times2$ ISD phase, as shown in Fig.~\ref{fig:schematic}.
To isolate the effects of $U$ and $V$ from structural contributions,
we use the optimized ISD structure obtained at $U=0$ and $V=0$
as the baseline for calculations with various $U$ and $V$ values.
The phase shifts between adjacent kagome layers
have minimal impact on the band dispersion.
Thus, our approach is sufficient for investigating the electronic structure of KV$_3$S$_5$.
Structural contributions,
such as variations in vanadium atom distortion for different values of $U$ and $V$,
will be discussed at the end of this subsection.

Figure~\ref{fig:bands} presents the band dispersion of the $2\times2$ ISD phase
for various values of $U$ and $V$,
unfolded onto the $1\times1$ Brillouin zone.
The $k$-path K-M-K is selected to facilitate direct comparison with
existing angle-resolved photoemission spectroscopy (ARPES) measurements,
as shown in Fig.~\ref{fig:bands}(e).

The band dispersion of the ISD phase
successfully captures two dominant features near the Fermi level:
(i) an M-shaped band and (ii) a $\Lambda$-shaped band,
as highlighted by the guidelines in Fig.~\ref{fig:bands}(a).
These features are consistent with the experimental ARPES measurement
by Kato \emph{et al.}~\cite{kato2022three},
where the former is denoted as the SP1 band and the latter as SP2.
Near the M-point, the $\Lambda$-shaped (SP2) band penetrates the M-shaped (SP1) band
and reaches a higher binding energy in the ISD phase.
In contrast, no such penetration occurs in the SD phase,
where the $\Lambda$-shaped (SP2) band remains at a higher energy position.

We found that our on-site and inter-site Hubbard parameters, $U$ = 1.35 eV and $V$ = 0.17 eV,
obtained from the cRPA calculations, successfully reproduce
the key features of the experimental ARPES
described above (Fig.~\ref{fig:bands} (a)).
Both the relative energy positions of the SP1 and SP2 bands
and their morphology closely match the experimental spectra in Fig.~\ref{fig:bands}(e).
This agreement contrasts with the results obtained using other $U$ and $V$ values,
as shown in the Fig.~\ref{fig:bands}.
While DFT calculations without $U$ and $V$ (Fig.~\ref{fig:bands}(b))
also show reasonable agreement with ARPES data,
the cRPA-derived finite $U$ and $V$ values play a crucial role
in shaping the band dispersion of the KV$_3$Sb$_5$ system,
as demonstrated in Fig. S5 of the Supplementary Material~\cite{Supp}.

Our systematic investigation reveals the distinct roles of $U$ and $V$
in shaping the low-energy band structures.
Variations in $U$ primarily affect the M-shaped (SP1) band,
which predominantly has
$3d_{x^2-y^2}$, $3d_{xy}$, and $3d_{z^2}$ orbital characters,
while the $\Lambda$-shaped (SP2) band,
mainly associated with $3d_{xz/yz}$ orbital characters,
remains nearly rigid
(see Fig. S7 in the Supplementary Material~\cite{Supp}
for detailed orbital characters).
Note that the orbital characters of both SP1 and SP2 bands
are consistent with the earlier findings by Kato \emph{et al.}~\cite{kato2022three}.
Increasing $U$ progressively shifts the SP1 band downward,
lowering its energy at the $K$-point,
whereas $U$ has minimal impact on the position of the SP2 band.
In contrast, the effect of $V$ is far more pronounced.
Fixing $U$ at 1.5 eV and varying $V$ from 0.5 to 1.5 eV,
we find that the low-energy band structure is highly sensitive to $V$.
As $V$ increases, the M-shaped nature of the SP1 band is completely disrupted,
and the relative position of the SP1 and SP2 bands changes significantly.

The band dispersion of the ISD phase
across the entire Brillouin zone
(calculated using the cRPA-derived values: $U$ = 1.35 eV and $V$ = 0.17 eV),
is provided in Fig. S6 \cite{Supp},
along with that of the pristine phase for comparison.

The positions of the van Hove singularities are essential
for identifying electronic instabilities
\cite{li2021observation,ferrari2022charge,xie2022electron}.
In the pristine phase, one of the prominent van Hove singularities appears near the M-point.
This feature is found to be split in the ISD phase, as demonstrated in Fig. S6 \cite{Supp}.
The split bands are associated with the M-shaped band shown in Fig.~\ref{fig:bands}(a).
The dominant orbital contributions to the M-shaped band
arise from the $3d_{x^2-y^2}$, $3d_{xy}$, and $3d_{z^2}$ orbitals,
which give rise to pronounced peaks near -0.2 eV
in the density of states (see Fig.~\ref{fig:DOS}).
Note that the van Hove singularity peaks are substantially suppressed
in the ISD phase relative to the pristine phase,
due to the band splitting near the M-point.

Now we account for structural contributions corresponding to
different $U$ and $V$ values, as discussed in Sec.~\ref{sub:stability},
which were neglected in the previous calculations.
Figure S5 (in the Supplementary Material~\cite{Supp}) shows
the band-unfolding results for the optimized $2 \times 2$ ISD structures
with corresponding $U$ and $V$ values.
For higher $V$ values, structural optimization induces
large atomic displacements of vanadium atoms, resulting in significant energy band splitting.
Notably, in the absence of $V$, the M-shaped nature of the SP1 band
is almost entirely suppressed as $U$ increases from 0 to 2 eV.
This suggests that the unfolded band dispersion closely resembles
that of the pristine structure (shown in Fig. S4~\cite{Supp}),
reinforcing our earlier analysis that CDW deformations diminish
and the system transitions towards a stable pristine structure as $U$ increases.
It is worth noting that the primary effect of the structural contributions
is the modification of the band dispersion arising from the larger atomic distortions
at finite $U$ and $V$ values.

\subsection{Electronic structures: density of states}

\begin{figure}[!t]
	\begin{center}
		\includegraphics[clip=true,scale=0.43]{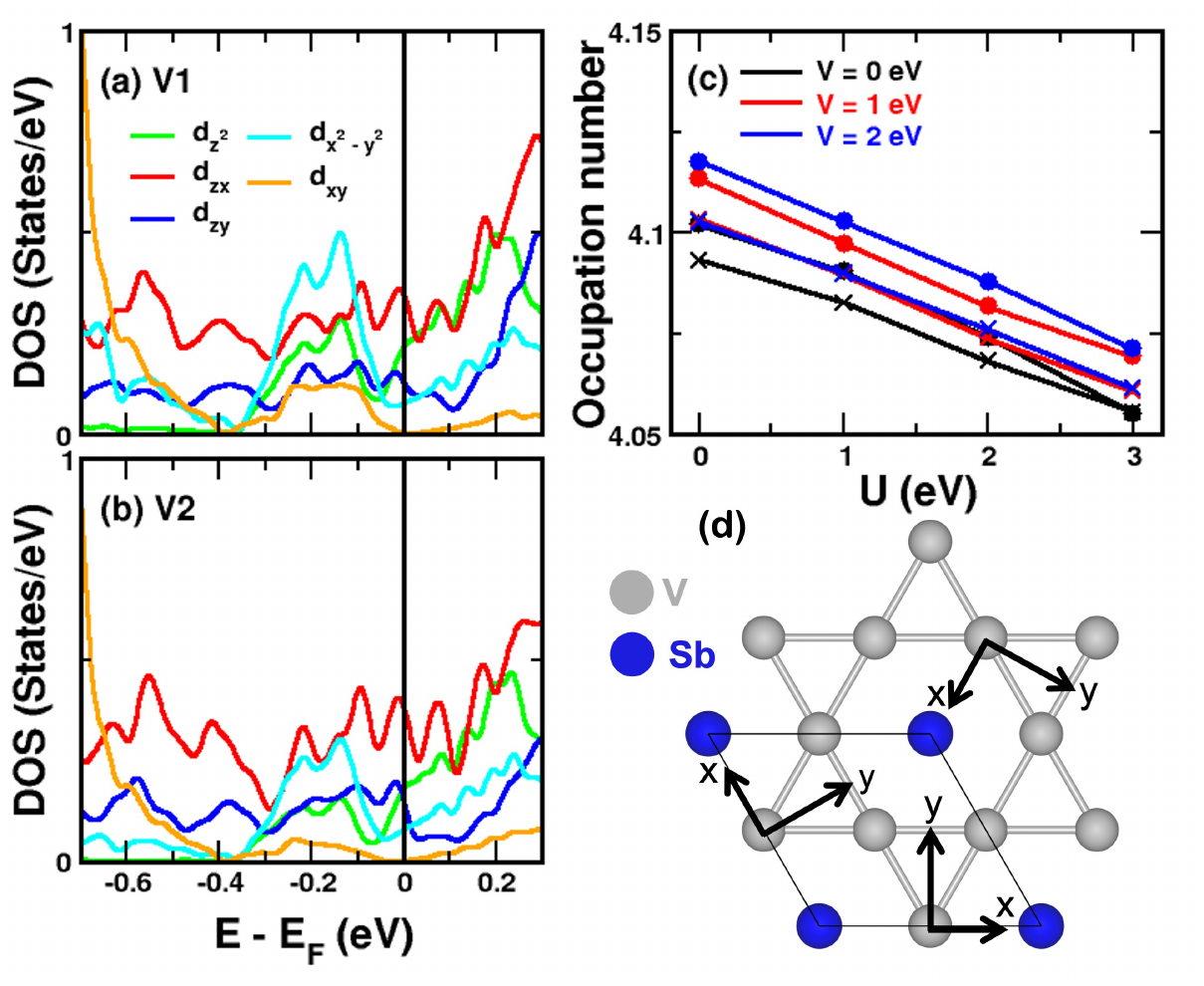}
	\end{center}
	\caption{(a) and (b) Orbital-resolved density of states for two different vanadium sites near the Fermi level in the $2\times2$ ISD phase, calculated using $U$ and $V$ obtained from cRPA. The Fermi level is set to 0 eV. (c) Calculated V-$3d$ occupation numbers for the $2\times2$ ISD phase obtained for different $U$ and $V$ values. The filled circles and cross symbols represent the two inequivalent vanadium sites V1 and V2, respectively. (d) Top view of the kagome lattice plane with its respective local (x, y) axes.
    The local x-axis points toward the Sb atom in the kagome lattice plane,
    while the local y-axis points to the center of the vanadium triangle.}
	\label{fig:DOS}
\end{figure} 

To gain deeper insight into the electronic structure of KV$_3$Sb$_5$,
we calculated the density of states (DOS) based on the local axis,
as shown in Fig.~\ref{fig:DOS}(d).
In the $2\times2$ ISD phase, vanadium atoms occupy two inequivalent sites
in the kagome lattice, denoted as V1 and V2 (Fig.~\ref{fig:schematic}).
The orbital-resolved DOS for V1 and V2 in the $2\times2$ ISD phase
is shown in Fig.~\ref{fig:DOS}.
To examine the effects of $U$ and $V$ solely,
we used the optimized ISD structure obtained at $U=0$ and $V=0$.
The DOS for V1 and V2 exhibit similar behavior,
with V1 contributing slightly more than V2 near the Fermi level,
which remains unchanged for all values of $U$ and $V$.
Furthermore, in the absence of $V$, increasing $U$ enhances the DOS at the Fermi level,
whereas incorporating $V$ (without $U$) leads to the opposite trend.
Specifically, $V$ suppresses the DOS at the Fermi level,
except for the $d_{xz/yz}$ orbitals,
which contribute to the $\Lambda$-shaped (SP2) band,
as shown in the projected band dispersion in Fig. S7~\cite{Supp}.
This interplay between $U$ and $V$ in shaping the PDOS at the Fermi level
is illustrated in Fig. S8 of the Supplementary Material~\cite{Supp}.
It is well established that CDW phases disrupt van Hove singularities,
leading to a reduction in the DOS at the Fermi level~\cite{tan2021charge}.
These finding align with our earlier results in Sec.~\ref{sub:stability},
where we observed that increasing $U$ destabilizes the CDW phase,
making the pristine structure more favorable,
while higher values of $V$ help sustain the CDW phase.

We further calculated the $3d$ occupation numbers for V1 and V2
across different values of $U$ and $V$.
The occupation numbers were determined by integrating the partial DOS of the $d$-orbitals up to the Fermi energy.
Figure~\ref{fig:DOS}(c) shows the occupation numbers for V1 and V2 as a function of $U$ for different $V$ values in the $2\times2$ ISD phase.
For $U=0$ and $V=0$, the occupation number of V1 (V2) is nearly 4.102 (4.093),
decreasing to 4.055 (4.056) at $U=3$ eV and $V=0$.
Our results confirm that for any given $V$,
the occupation number decreases with increasing $U$,
whereas for a fixed $U$, increasing $V$ leads to the opposite trend.
The orbital-decomposed occupation numbers, shown in Fig. S9~\cite{Supp},
further illustrate this contrasting behavior of $U$ and $V$.

\section{\label{sec:level5}Conclusions}
In summary, we systematically investigated the roles of on-site ($U$) and inter-site ($V$) Coulomb interactions in shaping the charge-density-wave (CDW) phases of the kagome metal KV$_3$Sb$_5$.
From our DFT+$U$+$V$ calculations, we constructed the $U$-$V$ phase diagram, revealing the contrasting roles of $U$ and $V$ in forming CDW phases. We found that $V$ plays a decisive role in stabilizing the inverse Star of David (ISD) CDW phase, consistent with experimental observations. The interplay between $U$ and $V$ influences not only the relative stability and extent of lattice distortions but also modulates the electronic instabilities associated with van Hove singularities near the Fermi level. By examining unfolded band structures, we clearly demonstrated how variations in $U$ and $V$ modify the band shapes and dispersions. Our cRPA estimates provided realistic values of $U$ and $V$, enabling us to identify the position of KV$_3$Sb$_5$ in the $U$-$V$ phase space. Through comparison with experimental ARPES data, we demonstrated that our chosen interaction parameters accurately reproduce the band structure of the system.

Our findings underscore the importance of both local and nonlocal
Coulomb interactions in kagome systems and provide a unified framework for understanding the emergence of exotic phases—including unconventional superconductivity and topological phenomena—in kagome metals. This work offers valuable insights for future studies of correlated states in kagome-lattice materials.

\section*{Acknowledgements}
BK acknowledges support from NRF (NRF2021R1A4A1031920, RS-2021-NR061400, and RS2022-NR068223) and KISTI Supercomputing Center
(Project No. KSC-2023-CRE-0413). CJK was supported by the National Research Foundation of Korea (NRF) grant funded by the Korean Government (MSIT) (Grant No. 2022R1C1C1008200). The authors acknowledge the support from the Advanced Study Group
program from PCS-IBS and the hospitality at APCTP, where part of this work was done.

\bibliography{ref}

\end{document}


\preprint{APS/123-QED}

\title{Supplementary Material of \\ ``Role of On-site and Inter-site Coulomb Interactions in KV$_3$Sb$_5$: A first-principles DFT+$U$+$V$ study''}

\author{Indukuru Ramesh Reddy}
\thanks{These authors contributed equally to this work}
\affiliation{Department of Physics, Kyungpook National University, Daegu 41566, Republic of Korea}
\affiliation{Department of Physics, Kunsan National University, Gunsan 54150, Republic of Korea}
\author{Sayandeep Ghosh}
\thanks{These authors contributed equally to this work}
\affiliation{Department of Physics, Kunsan National University, Gunsan 54150, Republic of Korea}
\affiliation{Department of Physics, Chungnam National University, Daejeon 34134, Republic of Korea}
\author{Bongjae Kim}
\email{bongjae@knu.ac.kr}
\affiliation{Department of Physics, Kyungpook National University, Daegu 41566, Republic of Korea}
\affiliation{Department of Physics, Kunsan National University, Gunsan 54150, Republic of Korea}
\author{Chang-Jong Kang}
\email{cjkang87@cnu.ac.kr}
\affiliation{Department of Physics, Chungnam National University, Daejeon 34134, Republic of Korea}
\affiliation{Institute of Quantum Systems, Chungnam National University, Daejeon 34134, Republic of Korea}

\date{\today}


\maketitle


\onecolumngrid


In KV$_3$Sb$_5$, there are eight symmetry-independent Wyckoff positions in $2\times2\times1$ ISD and SD phase: K1 (0.5, 0, 0), K2 (0, 0, 0), V1 ($x_1$, 0, 0.5),
V2 ($x_2$, 2$x_2$, 0.5), Sb1 ($x_3$, 2$x_3$, $z_1$), Sb2 (1/3, 2/3, $z_2$), Sb3 (0.5, 0, 0.5), Sb4 (0, 0, 0.5). 
The independent coordinates for the ISD and SD phases, with and without considering $U$ and $V$ are listed in Table S1.

\setcounter{table}{0}
\renewcommand{\tablename}{Table.}
\renewcommand{\thetable}{S\arabic{table}}
\begin{table*}[h!!]
\begin{center}
\caption{Wyckoff positions of star of David (SD) and inverse star of David (ISD) CDW phase of KV$_3$Sb$_5$ with and without considering $U$ and $V$.}

\vspace{0.2cm}
\setlength{\tabcolsep}{8pt}
\renewcommand{\arraystretch}{1.7}
\begin{tabular}{  c  c  c |  c  c | c  c   }
\hline
\hline
&  \multicolumn{2}{c}{$U=0$, $V=0$} & \multicolumn{2}{c}{$U=0$, $V=1$} & \multicolumn{2}{c}{$U=1, V=0$} \\
\cline{2-7}
& ISD & SD & ISD & SD & ISD & SD \\
\cline{2-7}
$x_1$ & 0.24556 & 0.25360 & 0.24315 & 0.25621 & 0.24603 & 0.25324 \\
$x_2$ & 0.25401 & 0.24746 & 0.25662 & 0.24583 & 0.253152 & 0.24776 \\
$x_3$ & 0.83419 & 0.83200 & 0.83475 & 0.83074 & 0.83404 & 0.83226 \\
$z_1$ & 0.25120 & 0.26003 & 0.24777 & 0.26094 & 0.25175 & 0.25923 \\
$z_2$ & 0.25744 & 0.25449 & 0.25645 & 0.25221 & 0.257114 & 0.25453 \\

\hline
\hline
\end{tabular}
\label{tab:table1}
\end{center}
\end{table*}

\begin{table}[!!h]
	\centering
	\begin{center}
		\caption{On-site ($U \& J$), inter-site ($V \& J^{'}$), and effective ($U_{eff}=U-J$ \& $V_{eff}=V-J^{'}$) screened interaction parameters for the pristine phase of KV$_3$Sb$_5$ experimental structure, calculated using constrained random-phase approximation (cRPA). All values are given in eV.}
		\vspace{0.2cm}
		\setlength{\tabcolsep}{8pt}
		\renewcommand{\arraystretch}{1.7}
		\begin{tabular}{ c c | c c | c c}
			\hline
			\hline
			\multicolumn{2}{c|}{on-site} & \multicolumn{2}{c|}{inter-site} & \multicolumn{2}{c}{effective parameters} \\
			\cline{1-6}
			$U$ & $J$ & $V$ & $J^{'}$ & $U_{eff}$ & $V_{eff}$ \\
			\cline{1-6}
			1.776 & 0.426 & 0.175 & 0.006 & 1.35 & 0.17 \\ 	
			\hline
			\hline
		\end{tabular}
		\label{tab:table2}
	\end{center}
\end{table}

\setcounter{figure}{0}
\renewcommand{\figurename}{Fig.}
\renewcommand{\thefigure}{S\arabic{figure}}
\begin{figure}[!!t]
	\begin{center}
		\centering
		\includegraphics[scale=0.45]{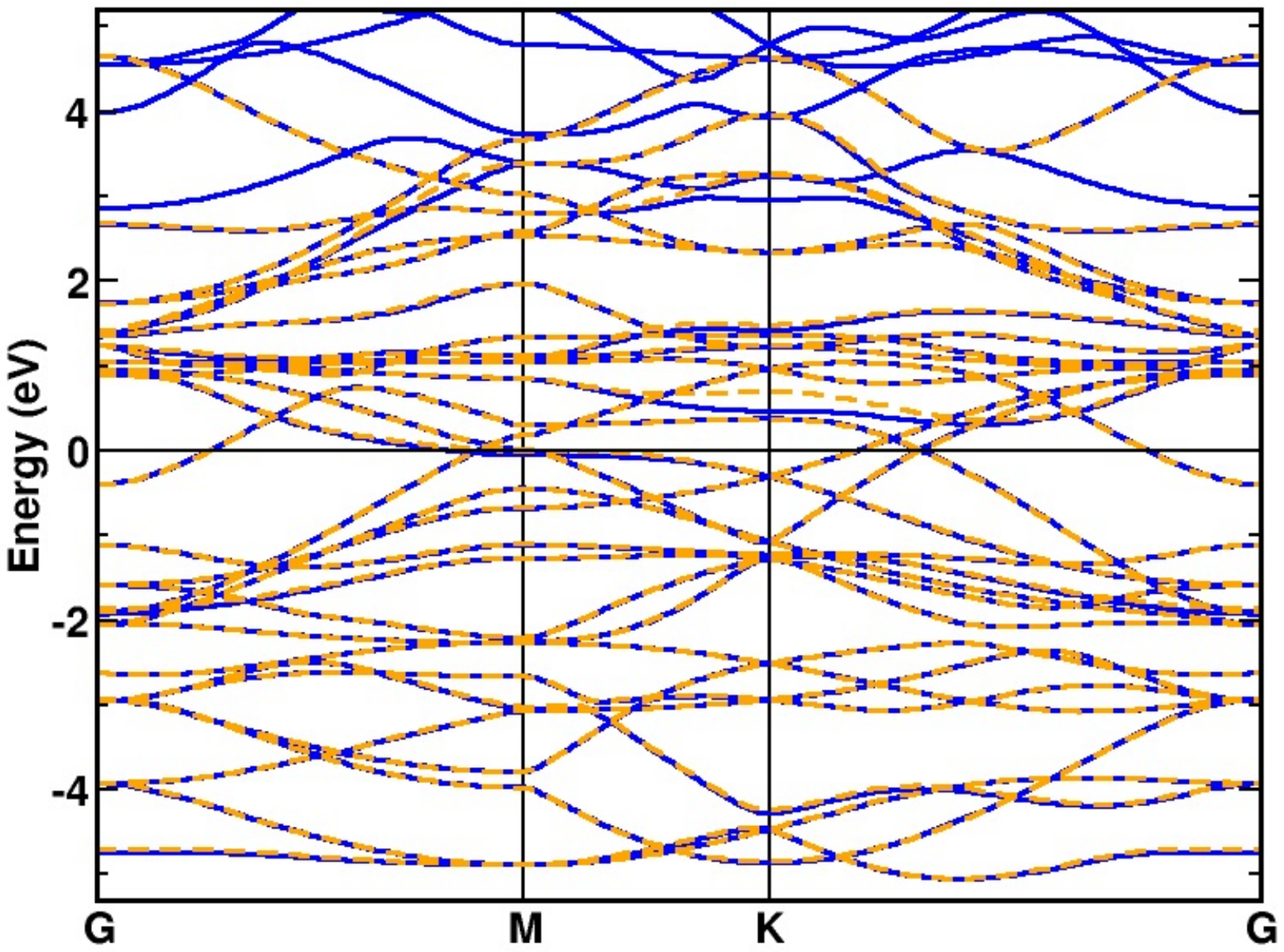}
	\end{center}
	\caption{Electronic structure of KV$_{3}$Sb$_{5}$ along with their Wannier-projected bands. $Ab~initio$ bands and Wannier-projected bands are represented by solid blue and dashed orange lines, respectively.}
	\label{fig:bs-wan}
\end{figure} 


\renewcommand{\figurename}{Fig.}
\renewcommand{\thefigure}{S\arabic{figure}}
\begin{figure}[!!b]
\begin{center}
 \includegraphics[trim=0.1cm 4.5cm 0.1cm 4.5cm, clip=true,scale=0.7]{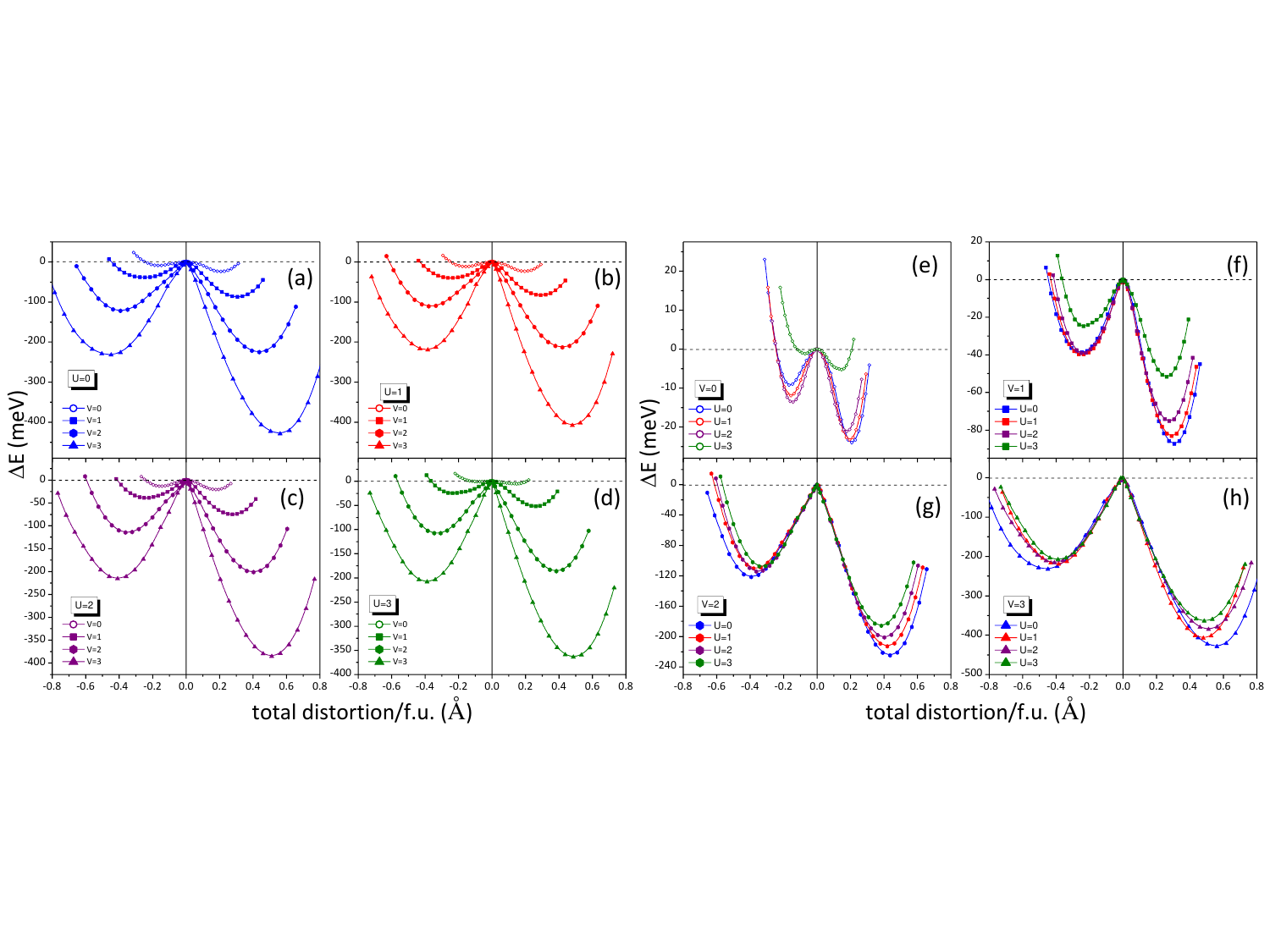}
\end{center}
\caption{(a) Variation of energy profiles as a function of distortion for CsV$_3$Sb$_5$: (a-d) for fixed $U =$ 0, 1, 2, and 3 eV with different $V$ and (e-h) for fixed $V =$ 0, 1, 2, and 3 eV with different $U$. The $\Delta$E represents the relative total energy concerning the pristine phase per supercell. The positive (negative) distortion attributed to the magnitude of the breathing-phonon mode of $2\times2$ CDW inverse star of David (star of David) phase.}
\label{fig:energy-profile}
\end{figure} 

\renewcommand{\figurename}{Fig.}
\renewcommand{\thefigure}{S\arabic{figure}}
\begin{figure}[!!t]
\begin{center}
 \includegraphics[trim=0cm 5.8cm 0cm 3.5cm, clip=true,scale=0.6]{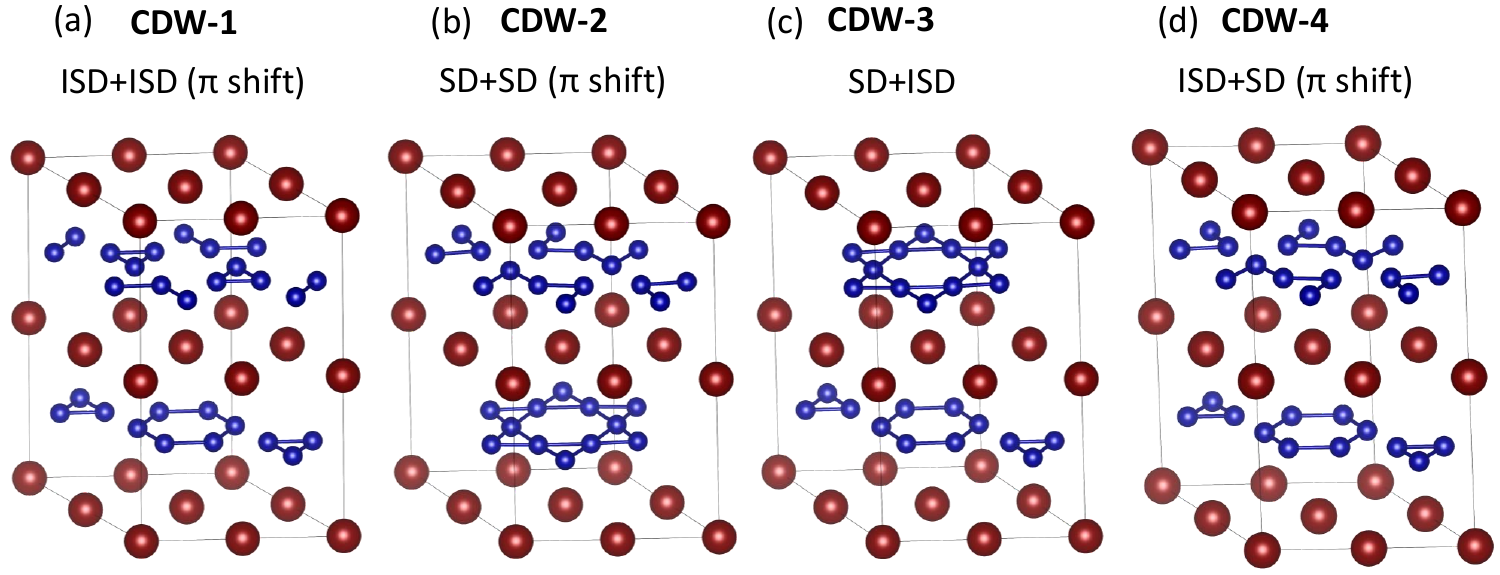}
\end{center}
\caption{Four different $2\times2\times2$ CDW structures of KV$_3$Sb$_5$ are constructed by combining the SD and ISD phase between the adjacent K layers: (a) CDW-1 (ISD+ISD with $\pi$-phase shift), (b) CDW-2 (SD+SD with $\pi$-phase shift), (c) CDW-3 (SD+ISD with zero phase shift), and (d) CDW-4 (SD+ISD with $\pi$-phase shift).
The ISD+ISD and SD+SD cases with zero phase shift corresponds to the $2\times2\times1$ ISD and SD CDW respectively. Sb atoms are removed for better visualization.}
\label{fig:CDW-222}
\end{figure} 

\renewcommand{\figurename}{Fig.}
\renewcommand{\thefigure}{S\arabic{figure}}
\begin{figure}[!!b]
\begin{center}
 \includegraphics[trim=4cm 1cm 4cm 0.5cm, clip=true,scale=0.7]{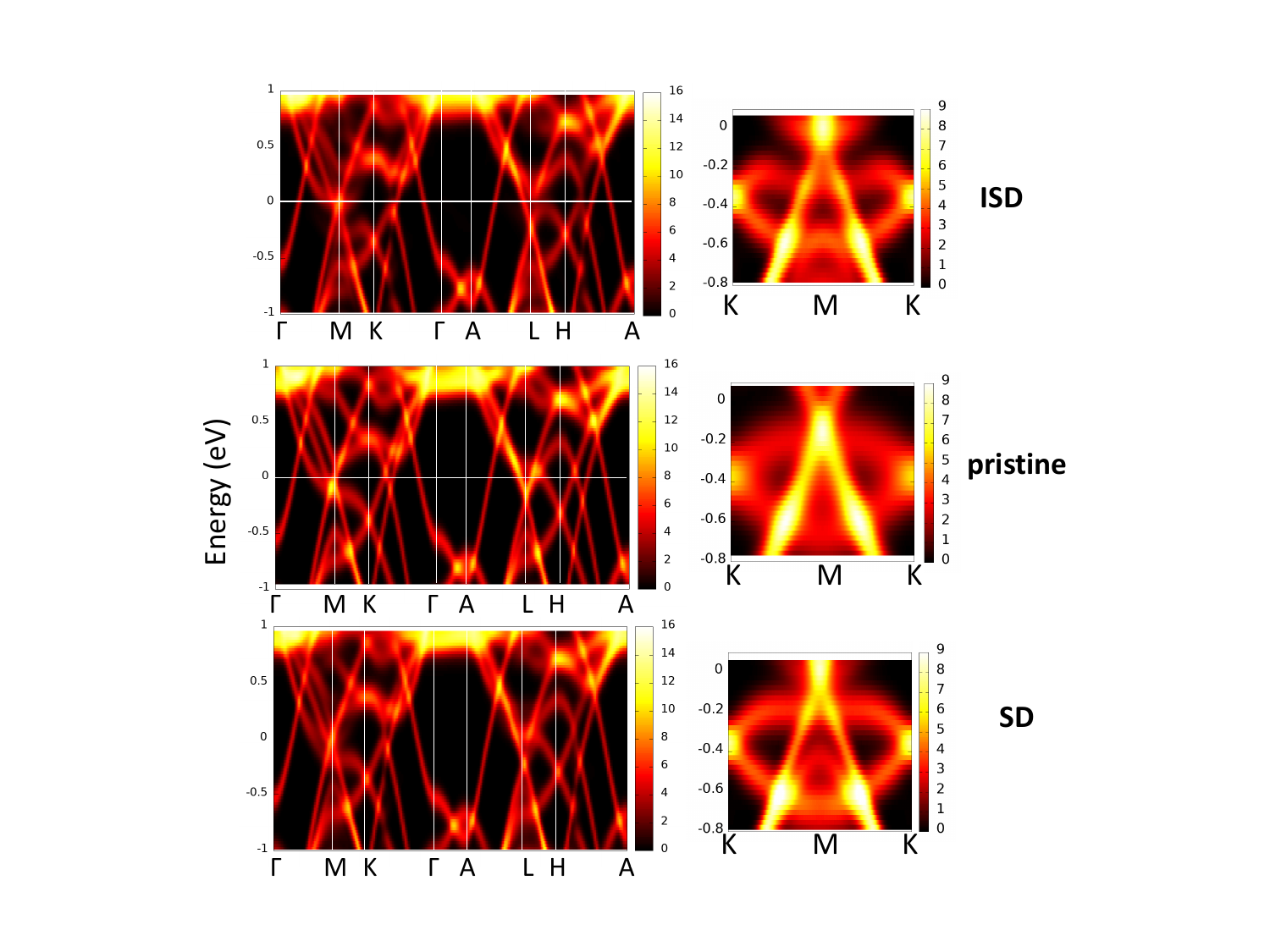}
\end{center}
\caption{Calculated unfolded band dispersion along the diiferent symmetry path for ISD, pristine and SD phases for $2\times2\times1$ supercell of KV$_3$Sb$_5$. The energy spectra are shown for $U=0$ eV and $V=0$ eV.}
\label{fig:schematic}
\end{figure} 

\renewcommand{\figurename}{Fig.}
\renewcommand{\thefigure}{S\arabic{figure}}
\begin{figure}[!!t]
\begin{center}
 \includegraphics[trim=2cm 1cm 1cm 0.5cm, clip=true,scale=0.9]{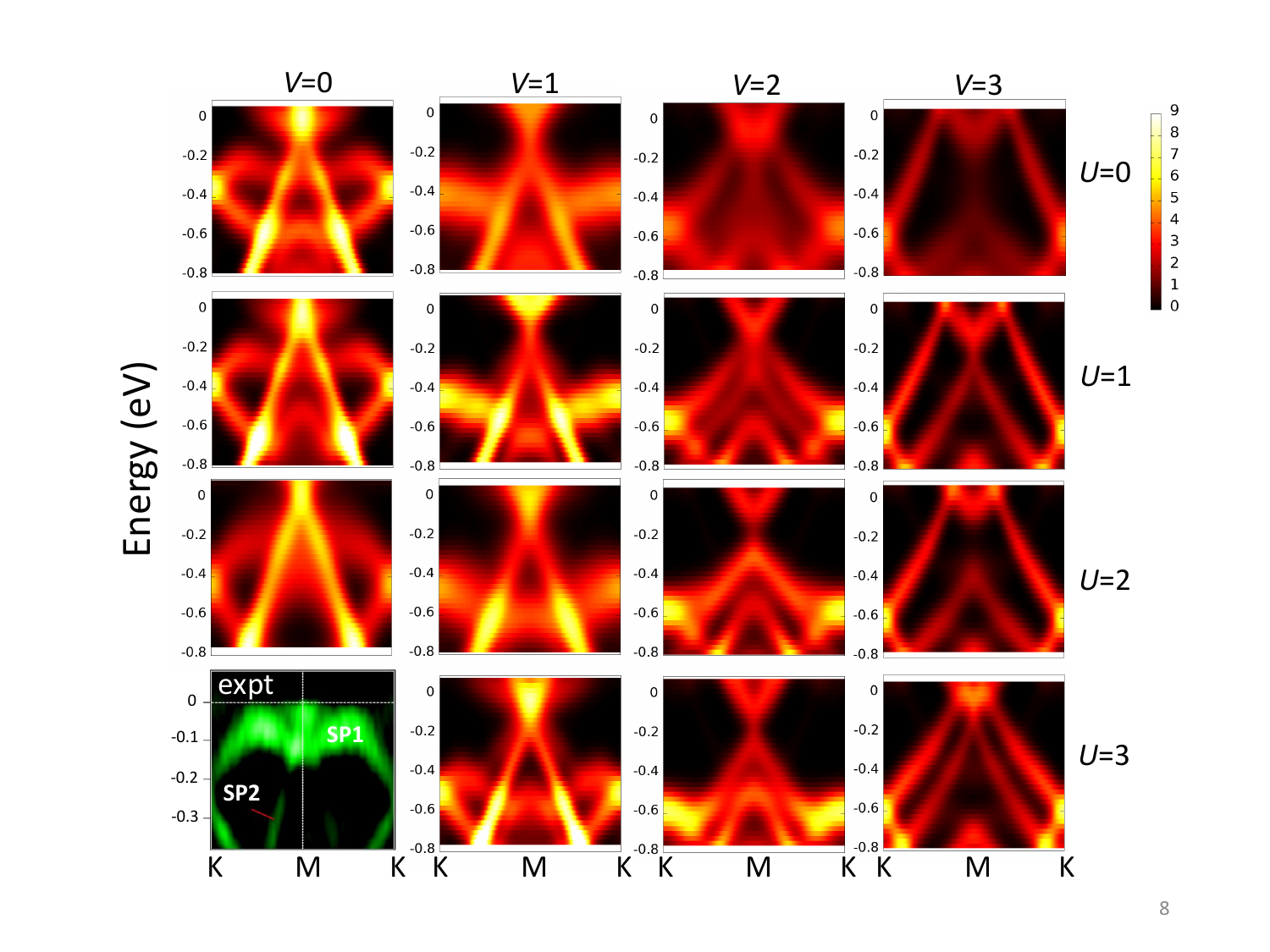}
\end{center}
\caption{Calculated unfolded band dispersion for optimized ISD structures for $2\times2\times1$ supercell of KV$_3$Sb$_5$ which is discussed in Sec. IIIB in the manuscript.}
\label{fig:unfolded-bs}
\end{figure} 

\renewcommand{\figurename}{Fig.}
\renewcommand{\thefigure}{S\arabic{figure}}
\begin{figure}[!!t]
\begin{center}
 \includegraphics[scale=0.7]{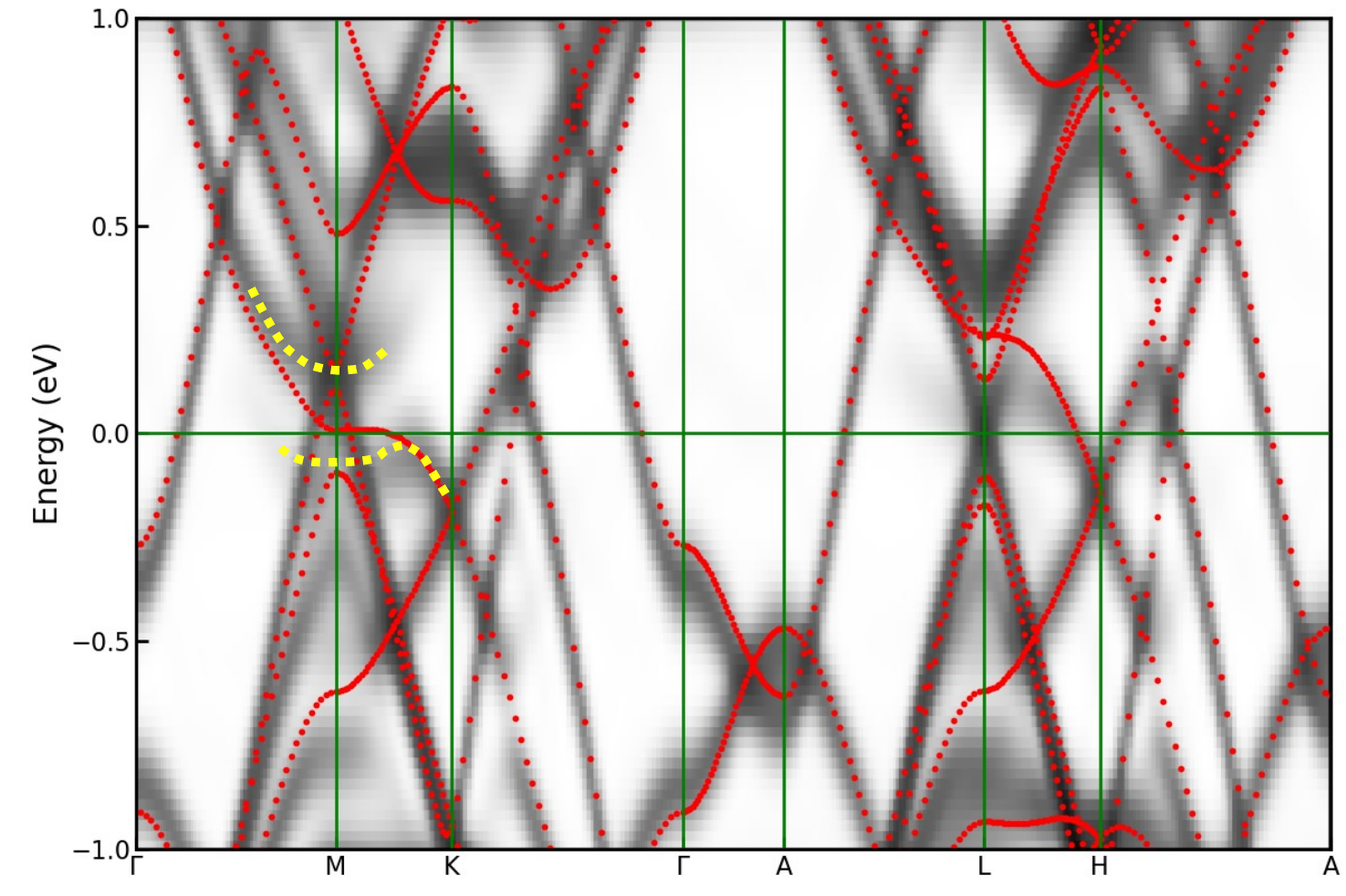}
\end{center}
\caption{Unfolded band dispersion for $2\times2$ ISD phase, calculated using $U$ = 1.35 eV and $V$ = 0.17 eV. The red dotted lines represents the bands for pristine phase (unit cell), obtained with $U$ = 1.35 eV and $V$ = 0.17 eV. The yellow dashed line is a guide to the eye, indicating the band associated with the van Hove singularity splitting at the M-point. The Fermi level has been shifted downward by 200 meV to be consistent with Fig. 4 in the main text.}
\label{fig:projected-bs}
\end{figure} 

\renewcommand{\figurename}{Fig.}
\renewcommand{\thefigure}{S\arabic{figure}}
\begin{figure}[!!t]
\begin{center}
 \includegraphics[trim=4cm 1cm 5cm 0.2cm, clip=true,scale=0.9]{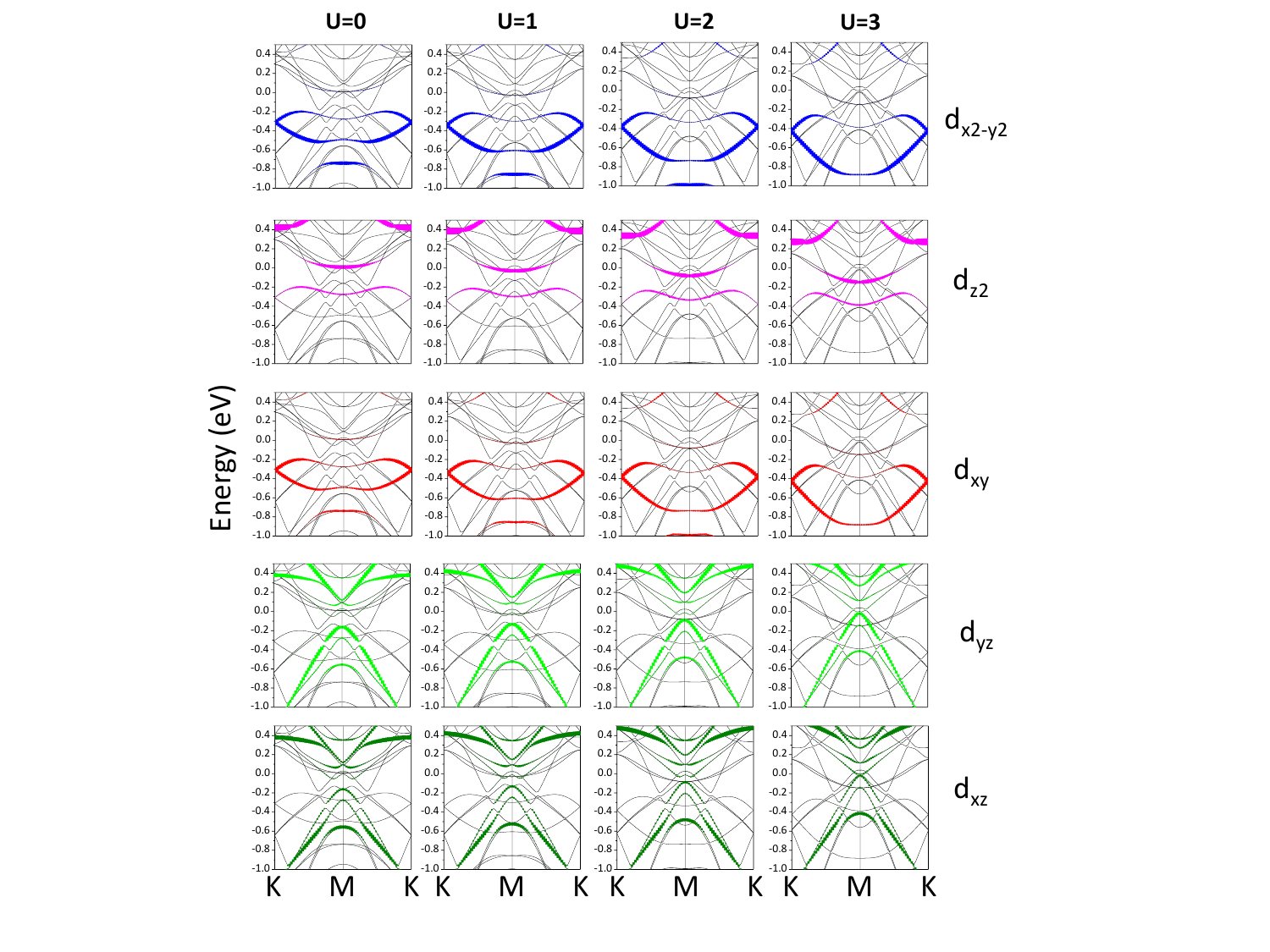}
\end{center}
\caption{Projected band dispersion of 3d orbital of vanadium atoms for $2\times2$ ISD deformation of KV$_3$Sb$_5$ for different $U$ and fixed $V$= 0 eV.}
\label{fig:projected-bs}
\end{figure} 

\renewcommand{\figurename}{Fig.}
\renewcommand{\thefigure}{S\arabic{figure}}
\begin{figure}[!!t]
	\begin{center}
		\includegraphics[clip=true,scale=0.65]{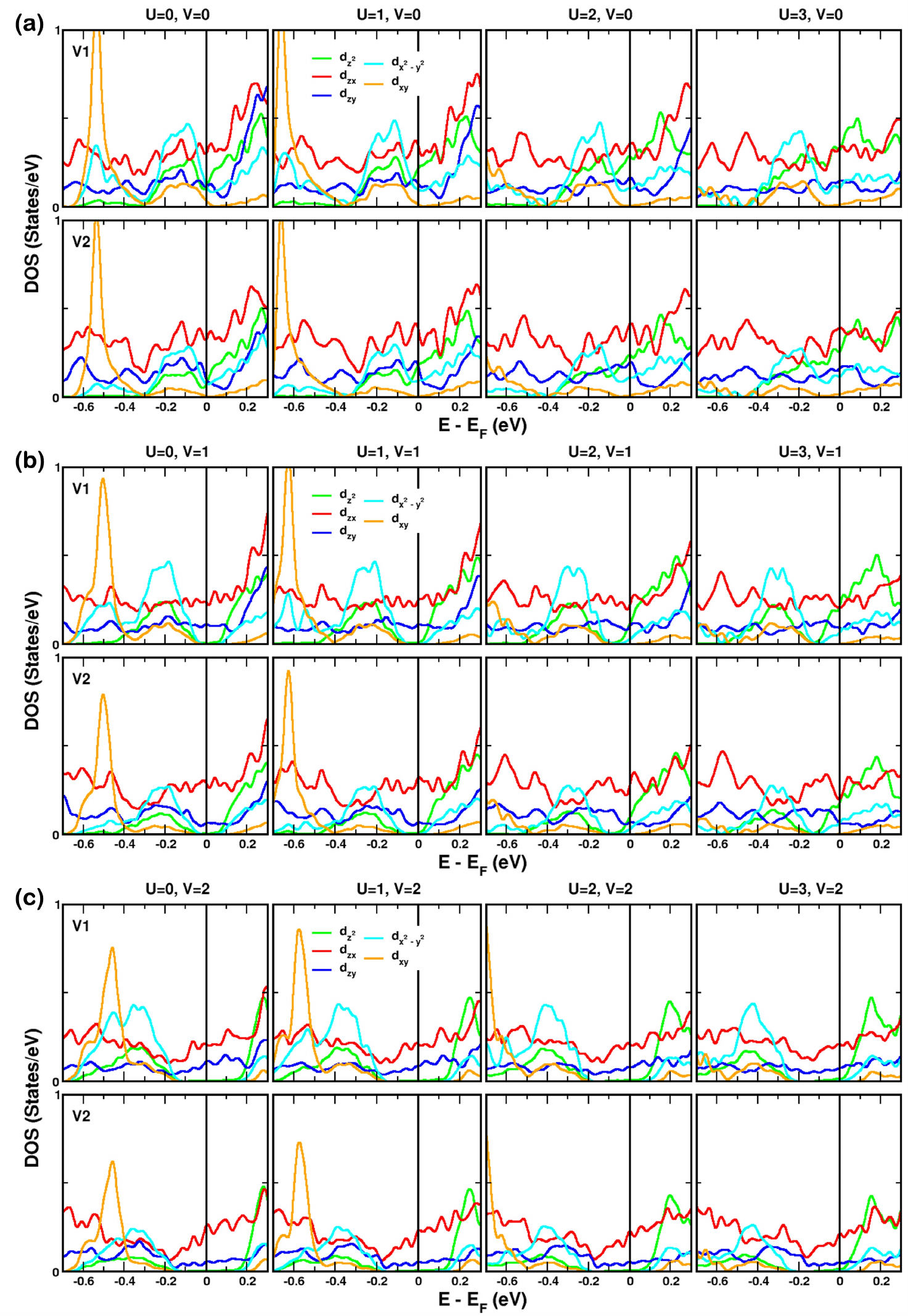}
	\end{center}
	\caption{Orbital-resolved DOS of two different vanadium sites near the Fermi level for the $2\times2$ ISD phase, calculated for different $U$ values from 0 to 3 eV, with fixed $V$: (a) $V$= 0 eV, (b) $V$ = 1 eV, and (c) $V$ = 2 eV. The Fermi level is shifted to 0 eV.}
	\label{fig:orb-dos}
\end{figure} 

\renewcommand{\figurename}{Fig.}
\renewcommand{\thefigure}{S\arabic{figure}}
\begin{figure}[!!t]
	\begin{center}
		\includegraphics[clip=true,scale=0.45]{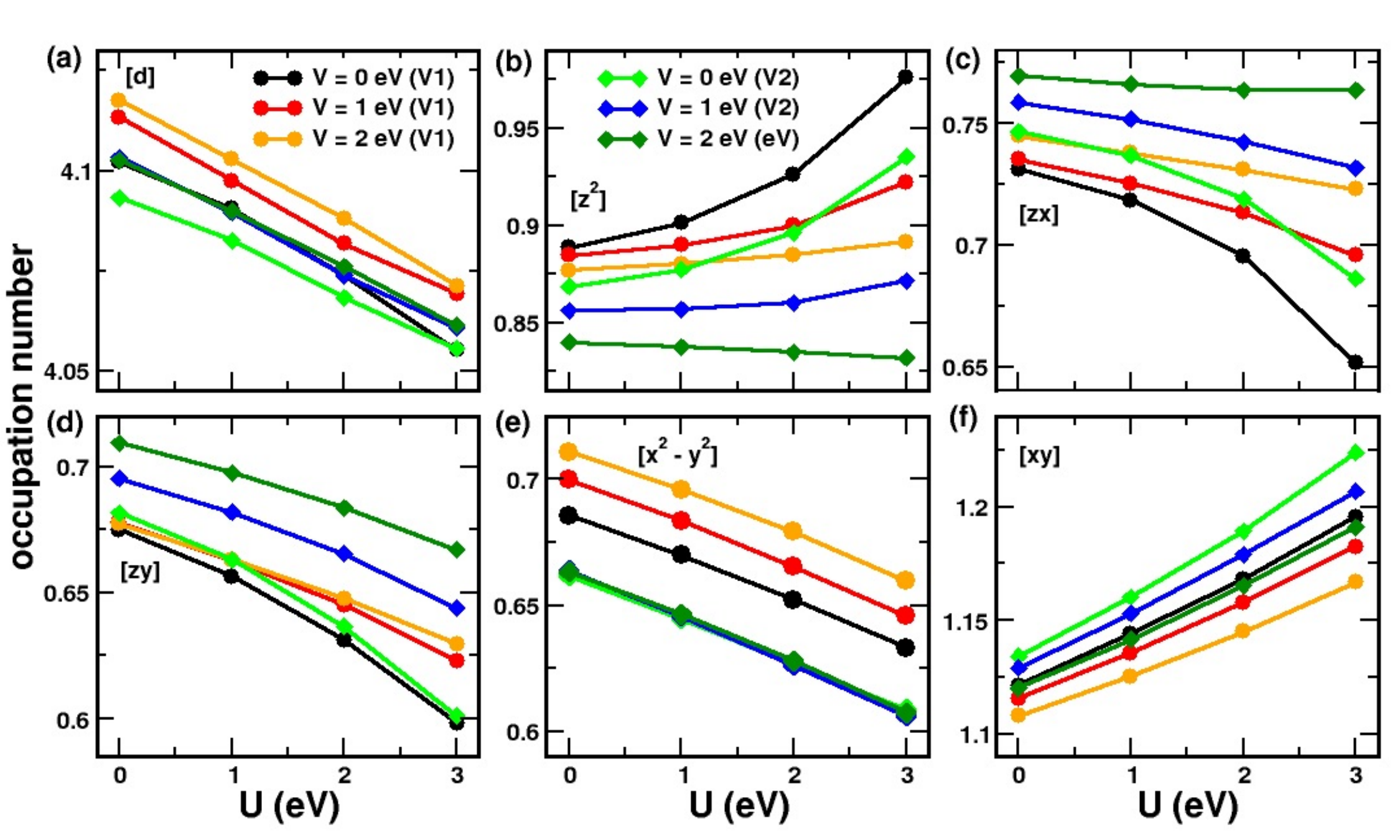}
	\end{center}
	\caption{Calculated occupation numbers for the $2\times2$ ISD phase  for different $U$ and $V$ values. (a) V-$d$, and (b)-(f) orbital-decomposed occupation numbers.}
	\label{fig:orb-occ}
\end{figure} 


